\documentclass[aps,prb,twocolumn,showpacs,superscriptaddress]{revtex4}
\usepackage{graphicx}
\usepackage{dcolumn}
\usepackage{bm}
\usepackage{color}

\begin{document}
\title{A quantitative study of spin-flip co-tunneling transport in a quantum dot}
\author{Tai-Min Liu} \affiliation{Department of Physics, University of Cincinnati, Cincinnati, OH 45221, USA}
\author{Anh T. Ngo} \affiliation{Department of Physics and Astronomy, Ohio University, Athens, OH 45701, USA}
\author{Bryan Hemingway} \affiliation{Department of Physics, University of Cincinnati, Cincinnati, OH 45221, USA}
\author{Steven Herbert}
\affiliation{Physics Department, Xavier University, Cincinnati, OH 45207, USA}
\author{Michael Melloch}
\affiliation{School of Electrical and Computer Engineering, Purdue University, West Lafayette, Indiana 47907, USA}
\author{Sergio E. Ulloa} \affiliation{Department of Physics and Astronomy, Ohio University, Athens, OH 45701, USA}
\author{Andrei Kogan}
\email{andrei.kogan@uc.edu}
\affiliation{Department of Physics, University of Cincinnati, Cincinnati, OH 45221, USA}
\date{\today}

\begin{abstract}
We report detailed transport measurements in a quantum dot in a spin-flip cotunneling regime and quantitatively compare the data to microscopic theory. The quantum dot is fabricated by lateral gating of a GaAs/AlGaAs heterostructure, and the conductance is measured in the presence of an in-plane Zeeman field. We focus on the ratio of the nonlinear conductance values at bias voltages exceeding the Zeeman threshold, a regime that permits a spin flip on the dot, to those below the Zeeman threshold, when the spin flip on the dot is energetically forbidden. The data obtained in three different odd-occupation dot states show good quantitative agreement with the theory with no adjustable parameters. We also compare the theoretical results to the predictions of a phenomenological form used previously for the analysis of non-linear cotunneling conductance, specifically in the determination of the heterostructure g factor, and find good agreement between the two approaches.  The ratio of nonlinear conductance values is found to slightly exceed the theoretically anticipated value and to be nearly independent of dot-lead tunneling coefficient and dot energy level.
\end{abstract}

\maketitle

 Electronic transport in nanoscale devices \cite{motayed:07,Koppens:06,Han:10,Xiao:04,Kroger:09,Ilani:10,Fuechsle:10} has been of significant recent interest, in part for its use as a spectroscopic tool for precision studies of fundamental phenomena, and because of the relevance of these devices to spintronics and quantum computation. \cite{Loss:98,Awschalom:02,Simmons:09} For spintronics, it is important to understand how the spin state of a nanosystem couples to its host surroundings. Spin-dependent transport can be conveniently studied in tunable quantum dots (QD)s. \cite{moore:00,costi:00,Kogan:04,zumbuhl:04,Amasha:05} Using a dot weakly coupled to the ``leads" with an applied in-plane magnetic field,   Kogan \textit{et al.} \cite{Kogan:04} showed that the differential conductance $G=dI/dV_{ds}$ exhibits steps at $V_{ds}$ values given by the ratio of the Zeeman energy and the electron charge, $e$, and used a phenomenological fit to the transport data to measure the heterostructure $g$ factor.\cite{Kogan:04,liu:09}  Later, Lehmann and Loss \cite{Loss:06} developed a microscopic theory to calculate the conductance through a QD in this regime, which included phonon-assisted spin-flip mechanisms. In this paper, we present extensive transport data of a quantum dot in the spin-flip  cotunneling regime and compare the results to microscopic theory.\cite{Loss:06}  Importantly, we measure all dot parameters needed for the calculation of the conductance, which enables a direct comparison between the data and the microscopic theory without any adjustable parameters, and find excellent quantitative agreement between the data and theory.
\begin{figure}
\includegraphics[width=3.4in,keepaspectratio=true]{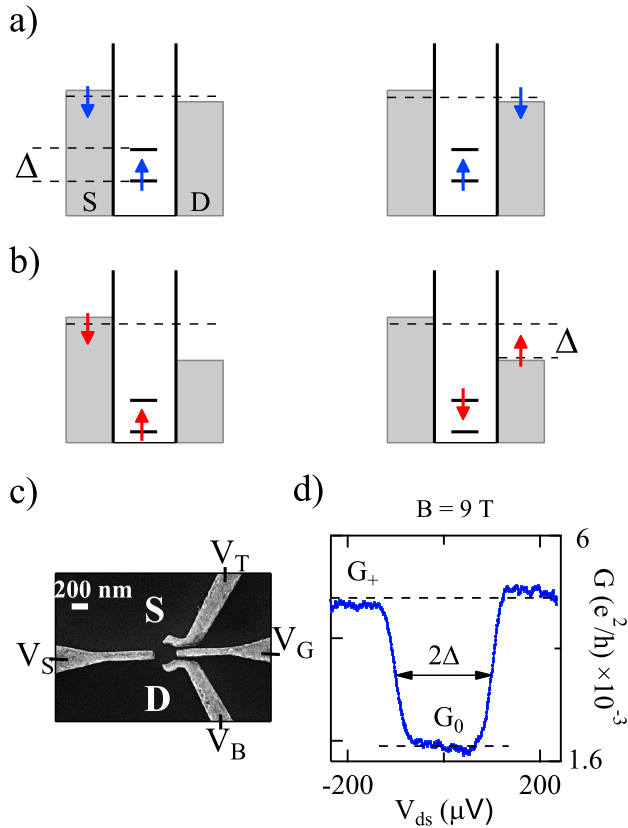}
\caption{\label{fig:1} (Color online) Cotunneling process through a Zeeman split orbital occurs when the dot is occupied by an odd number of electrons in the Coulomb blockade regime. Taking spin up to be the lower-energy state, a spin-down electron from the lead can tunnel onto and off the dot resulting in non-spin-flip cotunneling as shown in (a). When the bias voltage exceeds the Zeeman threshold, $|eV_{ds}| \geq \Delta=|g|\mu_B B$, the spin-up electron can also tunnel off the dot resulting in spin-flip cotunneling as shown in (b). (c) Micrograph of a device nominally identical to the one used in this paper. A quantum dot is created after applying negative voltages on electrodes $V_T$, $V_S$, $V_B$, and $V_G$. The electrode $V_S$ is used primarily to vary the dot-lead tunneling rate $\Gamma$ while the plunger gate $V_G$ is used to tune the dot energy. Differential conductance is measured through source(S) and drain(D) via standard lock-in techniques with 2 $\mu$V$_\mathrm{RMS}$ excitation at 17 Hz.  (d) Differential conductance as a function of drain-source voltage $V_{ds}$ in the cotunneling regime shows lower conductance ($G_0$) at $|eV_{ds}| < \Delta$ and higher conductance ($G_{+}$) at  $|eV_{ds}| \geq \Delta$. Dashed lines are guides for the average conductance values of $G_0$ and $G_+$.}
\end{figure}

We present data obtained for three different choices of the dot potential defined by the voltages on the confining gates, which correspond to three different occupancies of the dot. We focus on the ratio of the device conductance above and below the Zeeman threshold as a function of the tunneling rate and the dot energy. Since the orbital part of the wave function of the two Zeeman spin states is the same, the tunneling probabilities for each electron crossing the dot depend only on its spin and the spin of the dot. Therefore, a useful insight can be obtained from the ratio of the device conductance above and below the Zeeman threshold (i.e., when the bias across the dot matches the ratio of the Zeeman energy and the electron charge).  If the coupling to the leads is extremely weak (i.e., the tunneling rates between the dot and the leads are much smaller than the spin relaxation rate on the dot)  one might expect this ratio to be approximately 2:  at large biases, there are two possible dot states (the ground spin state and the excited spin state) available upon the completion of each tunneling event, whereas at low biases, the dot has to remain in the   ground spin state.  In practice, however, the spin relaxation rate due to intra dot processes is usually very slow, compared to the  tunneling rates in transport experiments between the dot and the leads.\cite{Hanson_prl:03, Amasha_prl:08} In that regime, therefore, exchanging spin with the leads is the dominant mechanism of the dot spin relaxation.  Predicting the device conductance in this regime requires a formalism that includes a complete set of rate equations, as we use in this paper for a single-orbital, spin 1/2 dot.\cite{Loss:06}  Our calculations and measurements both do reveal a nontrivial value for the conductance ratio $\approx 2.4$, indicating the important role of the current leads in providing spin relaxation in the dot.

Further, we show  that this ratio is independent of the dot-lead tunneling rate $\Gamma$ over approximately one decade,  $0.02<\Gamma<0.2$ meV, but varies slightly with the dot energy, exhibiting a slight minimum in the middle of the Coulomb blockade (CB) valley. Finally, we compare the shape of the nonlinear conductance as function of the dot bias as obtained from our calculations to the predictions of a phenomenological form used in earlier work. \cite{Kogan:04}  For a given Zeeman energy, we find excellent  agreement between the two, which means that either method provides a valid choice for using cotunneling transport for $g$-factor measurements. For the device used in this work, using both methods, we find the $g$ factor to be 0.2073 $\pm$ 0.0013.

The QD we have studied is created by gating a GaAs/AlGaAs heterostructure. Ti/Au electrodes of our single electron transistor (SET) are patterned via e-beam and photolithography followed by lift-off. The two-dimensional electron gas (2DEG) under the electrodes is statically depleted to form an electron droplet (i.e., a QD) connected on both sides to the electron reservoirs: source and drain [Fig.\ \ref{fig:1}(c)]. We estimate the diameter of the QD to be $\sim$ 0.13 $\mu$m, which contains tens of electrons. From magnetotransport data we find that the 2DEG has a mobility of $5 \times 10^5$ cm$^2$/(Vs) and an electron density of 4.8$\times$10$^{11}$cm$^{-2}$ at 4.2K. The  device is oriented parallel to the magnetic field within $\pm 1$ degree, and is cooled  in a Leiden Cryogenics dilution refrigerator to a base electron temperature $T_\mathrm{elec} \sim 55$mK. We use standard lock-in techniques to measure the differential conductance through the QD.

 Figure \ref{fig:1}(d) shows the differential conductance steps at  source-drain voltages equal to the Zeeman energy of the dot. The tunneling between the dot and the leads is relatively weak, so that the Kondo effect in this regime is suppressed by thermal fluctuations. In the Coulomb blockade (CB) regime, when the QD has an unpaired electron in the dot energy level, the spin degeneracy is removed by the Zeeman field, and the level splits into spin-up and spin-down states. We label the conductances below and above the Zeeman threshold as $G_0$ and  $G_+$, respectively. Figures \ref{fig:1}(a) and \ref{fig:1}(b) illustrate the possible tunneling processes: In an elastic event [Fig. \ref{fig:1}(a)], the dot is left in the ground state and the electron does not change its energy as it crosses the dot. If the dot is left in an excited state [Fig. \ref{fig:1}(b)], the electron energy is lowered by $\Delta$.

To examine the cotunneling conductance and the ratio of $G_+$ to $G_0$ quantitatively, we arrange three different dot configurations: COT I ($V_S=-800 \rightarrow -872$, $V_T=-816$, $V_B=-1151$, $V_G=-938 \rightarrow -792$ mV); COT II ($V_S=-960 \rightarrow -1025$, $V_T=-750$, $V_B=-1090$, $V_G=-795 \rightarrow -671$ mV); and COT III ($V_S=-800 \rightarrow -917$, $V_T=-750$, $V_B=-1090$, $V_G=-1246 \rightarrow -1008$ mV). For each configuration, the dot contains a different number of electrons.  To tune the dot-lead coupling, $\Gamma$, we use a previously developed computer control of the dot gate voltages \cite{liu:09} and adjust the voltages $V_S$ and $V_G$ so as to maintain the occupancy of the dot and keep the dot energy in the middle of the Coulomb valley. To tune the dot energy $|E_1-\mu|$, we vary the plunger gate voltage $V_G$ while keeping voltages on other electrodes unchanged.  We focus on the changes of the conductance as well as the ratio $G_+/G_0$ as either the tunneling rate or the dot energy is varied. The experiment is performed for all three device configurations described above.

\begin{figure}
\includegraphics[width=3.4in,keepaspectratio=true]{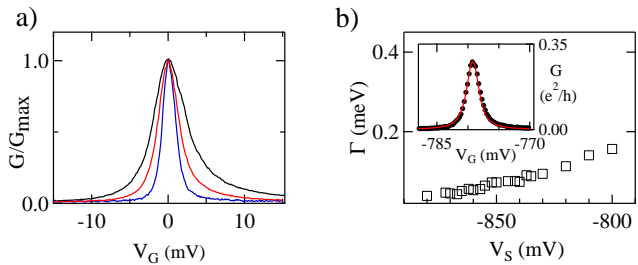}
\caption{\label{fig:2} (Color online) (a) Normalized Coulomb blockade peaks taken at $V_S$= -800, -830, and -872 mV (black, red, and blue lines respectively) for dot configuration I (COT I) clearly show the variation in peak width. (b) Tunneling rate $\Gamma$ as a function of $V_S$. Tunneling rate increases as the side gate voltage becomes less negative. Inset: Fitting a Coulomb blockade peak (dots) to a thermally broadened lorentzian (red solid line) gives the corresponding $\Gamma$. Configurations II and III show similar behavior (not shown).}
\end{figure}
\setcounter{table}{0}
\begin{table}
\caption[Capacitance ratio $\alpha_G$ for different $V_S$]{Capacitance ratio $\alpha_G$, used as the energy lever arm, extracted from Coulomb blockade diamonds for different $V_S$ values.  Tunneling rate $\Gamma$ and charging energy $U$ shown are parameters for the COT III dot configuration.}
\begin{tabular}{p{0.7in}p{0.7in}p{0.7in}p{0.7in}}
\\
\hline\hline
\
$V_S$ (mV) & $\alpha_G$ & $\Gamma$ (meV) & U (meV) \\
\hline
-800 & 0.027 & 0.19 & 2.77\\
-850 & 0.03  & 0.06  & 2.89\\
-900 & 0.036 & 0.03  & 3.11\\
\hline\hline
\end{tabular}
\end{table}

We measure the tunneling rate $\Gamma=\Gamma_L+\Gamma_R$, where $\Gamma_{L(R)}$ is the tunneling rate from the left(right) lead, by examining the shape of the charging peak as we vary the voltages on the gates.  Figure \ref{fig:2}(a) shows clearly the evolution of the Coulomb charging peak width as $V_S$ is varied. To determine $\Gamma$, we fit the CB conductance line shape to a thermally broadened lorentzian (TBL) \cite{foxman:93, foxman:94}
\begin{eqnarray}
\nonumber G(V_G)=\frac{e^2}{h}\frac{A}{4kT}\int_{-\infty}^{+\infty} \cosh^{-2}\left(\frac{E}{2kT}\right)\\
\times \frac{(\Gamma/2)\pi}{(\Gamma/2)^2+[e\alpha_G (V_G-V_{0})-E]^2}dE.
\end{eqnarray}
In this equation, $V_{0}$ is the  gate voltage that corresponds to the CB peak maximum, $\Gamma$ is the associated tunneling rate, $\alpha_G$ is the energy lever arm of the dot, and $A$ is a fitting parameter which is related to the dot asymmetry, \cite{Staring:92} $S=\Gamma_L/\Gamma_R$. The dot asymmetry for each $V_S$ voltage setting is obtained  from the height of the CB peak; \cite{Houten:05,stone:85} $S$ varies from 4 to 51 in our measurements. A slight deviation of $\alpha_G$ due to a possible shifting of the position of the dot has been observed and taken into consideration. Table I lists $\alpha_G$ and other dot parameters for three different choices of $V_S$. To assign the corresponding $\Gamma$ for each $V_S$, we use the average of the tunneling rates extracted from the two adjacent CB peaks in the same valley $\Gamma=(\Gamma_{LP}+\Gamma_{RP})/2$, where $\Gamma_{LP(RP)}$ corresponds to the left(right) CB peak. An approximately linear dependence of $\Gamma$ on $V_S$, and the TBL fitting to a CB peak are shown in Fig.\ \ref{fig:2}(b).  The overall conductance decreases with more negative $V_S$ values, as  expected, because of the reduction in the transmission of the barriers.

\begin{figure}
\includegraphics[width=3.4in,keepaspectratio=true]{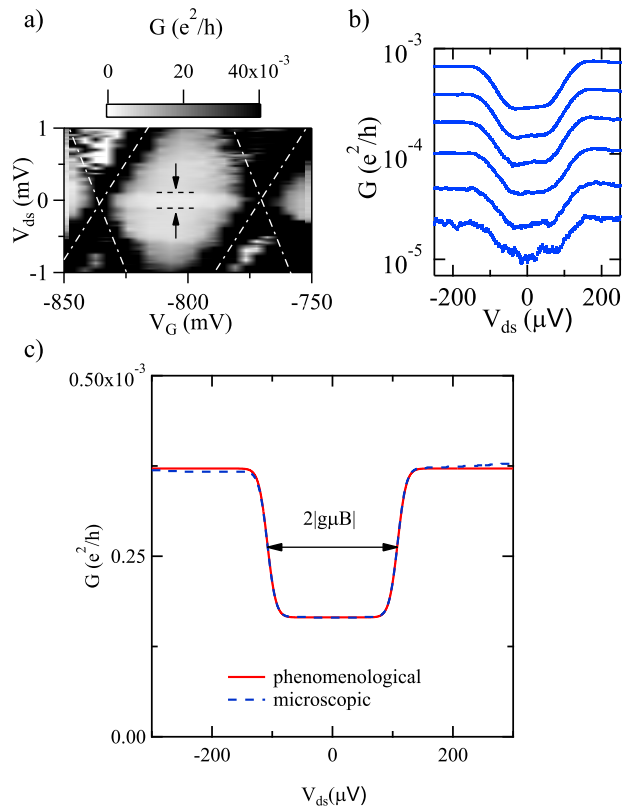}
\caption{\label{fig:3} (Color online) (a) Differential conductance as function of $V_{ds}$ and $V_G$ at $B=9$ T for COT III. The cotunneling trace of any given $\Gamma$ is taken in the middle of the Coulomb valley. Dashed lines with arrows indicate the conductance threshold across the odd-occupied valley. Dot-dashed lines mark the Coulomb blockade diamond edges.  (b) Representative differential conductance traces taken in the middle of Coulomb valley for different $\Gamma$s of COT II: from 0.072 meV (top) to 0.032 meV (bottom). (c) Comparison of the microscopic calculation with no adjustable parameters to predictions of a phenomenological form (Eq.\ 1 in Ref. [\onlinecite{Kogan:04}]) used for the data analysis shows good quantitative agreement. The gap width is twice the Zeeman energy.}
\end{figure}

Figure \ref{fig:3}(a) shows the characteristic features of the cotunneling conductance in the presence of the magnetic field, for a typical valley with an odd-number electron occupation. At each gate voltage, a threshold step is observed; the separation between the steps at positive and negative bias is controlled by the Zeeman energy, and it is thus independent of the gate voltage. Figure \ref{fig:3}(b) shows representative traces at several different $\Gamma$s while the dot energy is kept in the mid-point of the Coulomb valley as described above.

In order to make direct comparison between theory and experiment, we consider a model where transport occurs across a quantum dot contacted to two leads, in the presence of a spin-flip mechanism due to the coupling of the quantum dot to a phonon bath. The Hamiltonian of the system is described by \cite{Loss:06,loss0104,Schrieffer:66}
\begin{eqnarray}
H= H_0 + H_{tun} + H_{sp}
\label{H}
\end{eqnarray}
where $H_0$ stands for the Hamiltonian of the isolated dot, the ideal
leads, and the  free phonons,
 \begin{eqnarray}
H_0 =\sum_{\sigma} \varepsilon_{\sigma}n_{\sigma}+Un_{\uparrow}n_{\downarrow}+\sum_{lk\sigma}\varepsilon_{lk}n_{lk\sigma}+\sum_q\hbar\omega_q n_q;
\label{H0}
\end{eqnarray}
here, $n_\sigma$ ($n_{lk\sigma}$) is the number operator of the electron in the dot (leads) with spin $\sigma$ and $U$ is Coulomb interaction between two electrons in the dot with opposite spins.  The last term in Eq.~\ref{H0} describes the
free phonon bath with occupation numbers $n_q$ and energy  $\hbar\omega_q$.

The hybridization  between the dot and the leads is described by the
tunneling Hamiltonian
\begin{eqnarray}
H_{tun}=\sum_{lk\sigma} V_{lk}c^\dagger_{l
k\sigma}d_{\sigma}+\text{H.c.},
\end{eqnarray}
 where $c^\dagger_{lk\sigma}$ ($d_{\sigma}$)  is the fermionic creation (destruction) operator of the electron on the leads (dot). Here we have assumed that tunnel matrix elements, $V_{lk}$, are spin independent.
Finally, the spin-phonon interaction is modeled by
\begin{eqnarray}
H_{sp}=\sum_{q}({M_{qx}}\sigma_x+M_{qy}\sigma_y)(a^\dagger_{-q}+a_{q}),
\end{eqnarray}
where the bosonic operator $a^\dagger_{-q}$ ($a_{q}$) creates (destroys) a
phonon in the mode $q$; ${M_{qx}},{M_{qy}}$ are the spin-phonon coupling amplitudes,\cite{Loss:06} and $\sigma_x, \sigma_y$
are Pauli matrices.

To calculate the differential conductance
$G=dI/dV$, we derive the current which crosses the quantum dot from the left ($L$) to
the right ($R$) lead. The current through the dot can be expressed by\cite{daniela,loss0104}
\begin{eqnarray}
I_{LR}=e\sum_{\sigma \sigma'}W_{L\sigma',R\sigma}P_{\sigma}
\end{eqnarray}
where $e$ is electron charge, $W_{L\sigma',R\sigma}$ is the
transition rate for an electron tunneling from $L$ lead (spin $\sigma'$) into $R$
lead (spin $\sigma$), and can in principle take into account elastic, inelastic, as well as
phonon-assisted elastic cotunneling processes;
 $P_{\sigma}$ is the occupation number of electrons in the dot which is governed by a master
equation\cite{Loss:06}
\begin{eqnarray}
\frac{dP_{\sigma}}{dt}=-\gamma_{\bar{\sigma}\sigma}P_{\sigma}+\gamma_{\sigma\bar{\sigma}}P_{\bar{\sigma}} \, ,
\end{eqnarray}
with the rate
$\gamma_{\bar{\sigma}\sigma}$
including spin-flip and inelastic cotunneling processes with a current lead, and spin-flip processes
due to the spin-phonon coupling. In the stationary limit, the solution of the master equation is
given by $P_{\sigma}=[1+\gamma_{\bar{\sigma}\sigma}/\gamma_{\sigma\bar{\sigma}}]^{-1}$. Detailed expressions and discussions for the different rates are found in the literature. \cite{Loss:06,loss0104}

\begin{figure}
\includegraphics[width=3.4in,keepaspectratio=true]{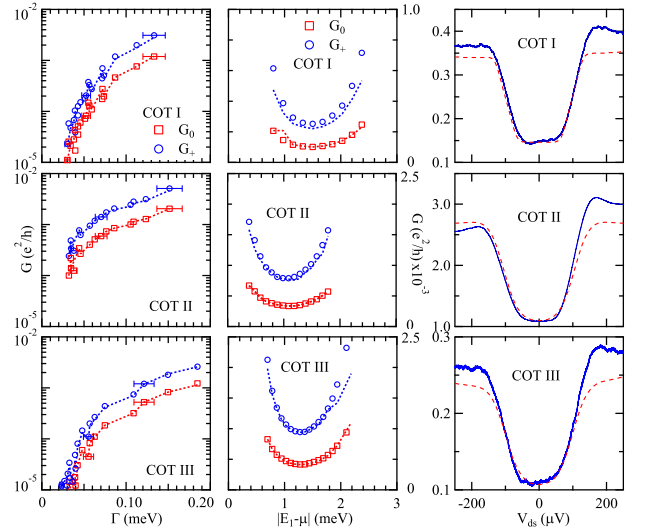}
\caption[Energy dependent result]{\label{fig:4} (Color online) Comparison of microscopic calculations (dotted lines) and measurements for $G_0$ (squares) and $G_+$ (circles) of three dot configurations. Left panels: dependence on the tunneling rate. Middle panels: dependence on the dot energy. Right panels: microscopic calculation (dotted line) of differential conductance as a function of $V_{ds}$ shows agreement with the experimental measurement (solid line) for the conductance near zero bias, but is slightly off at high bias.}
\end{figure}

Direct comparison of the calculated and experimental conductance traces, such as those in Fig.\ \ref{fig:4} (right panels),
shows their excellent agreement. \cite{FOOT}
Above the threshold, the measured conductance exceeds slightly the
calculated conductance, arising perhaps from a slight bias dependence in the barrier transmission coefficients and/or the increasing
important role of other dot levels ignored in the model.
We point out that spin-phonon interaction is expected to reduce the conductance at high bias; moreover, the
overshoot seen in the data near threshold is not expected for the strongly asymmetric quantum dots studied here.
\cite{Loss:06,Weymann:06}  Its nature is still unresolved.

We have also compared the microscopic theory to the phenomenological form used by Kogan \textit{et al.}\cite{Kogan:04}
for the analysis of non-linear cotunneling conductance.  We specifically use both approaches to determine the
$g$ factor of the heterostructure and find excellent agreement [Fig.\ \ref{fig:3}(c)] between both approaches.

\begin{figure}
\includegraphics[width=3.4in,keepaspectratio=true]{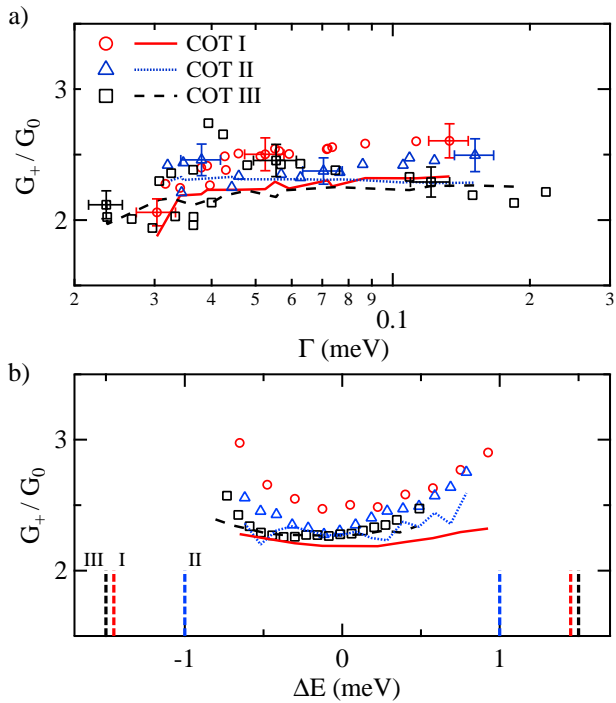}
\caption{\label{fig:5} (Color online) (a) The $G_+/G_0$ ratio as function of $\Gamma$. Both calculations (lines) and measurements (symbols) show that $G_+/G_0 \approx$ 2.4 is nearly independent of the tunneling rate $\Gamma$. (b) The conductance ratio as function of dot energy--using as reference the midpoint of the valley. Vertical dashed lines indicate where charging peaks appear (at half of the charging energies) for all three dot configurations. Charging energy values for COT I, II, and III are 2.9, 2.0, and 3.0 meV, respectively.  The ratio reveals a minimum  at the midpoint of the valley $\Delta E=0$, but it slightly increases as the dot energy approaches the charging peaks. }
\end{figure}

Having obtained the dot energy, $\Gamma$, $g$ factor, and the dot asymmetry,
we now focus on the conductances ($G_0$) and ($G_+$) for the three different dot configurations.
Figure \ref{fig:4} (left panels) shows quantitative agreement between the predictions and the data, for over two orders of magnitude in conductance, as $\Gamma$ changes.  Notice that the theoretical curves are not smooth functions of $\Gamma$ since the asymmetry factor is not the same for each choice of $\Gamma$. The ratio of conductances $G_+/G_0 \approx 2.4$, however, is nearly independent of  $\Gamma$ for all three configurations [Fig.\ \ref{fig:5}(a)].

Next, we address the variations of $G_0$ and $G_+$ with dot energy. The dot energy is tuned by varying the plunger gate voltage $V_G$, while maintaining $|E_1-\mu|/\Gamma$ or $(U-|E_1-\mu|)/\Gamma \ge$ 4, to avoid the dot entering the mix-valence regime. We find that the conductance increases symmetrically as the dot energy is tuned away from the mid-point of the valley [Fig.\ \ref{fig:4} (middle panels)]. We examine the ratio $G_+/G_0$ and find that although nearly constant at $\approx 2.4$, it exhibits a slight minimum at the midpoint of the valley; the ratio increases slightly as the dot energy approaches the adjacent CB peaks. Figure \ref{fig:5}(b) again shows good agreement between the calculated and measured results.

In summary, we have presented a systematic study of the differential conductance of a quantum dot in the cotunneling regime for three different dot occupancy configurations.  This allowed us to investigate the dependence of tunneling rate and dot energy on conductance and compare the experimental data to microscopic calculations. Independent experiments to determine the parameters of the dot state were performed so that comparisons could be made without use of adjustable parameters. We find overall excellent agreement between the calculations of a simple two-spin quantum dot model  and the measurements. We find that the ratio of the device conductance above the Zeeman threshold to that below the threshold is nearly independent of the dot-lead tunneling rate and it is only slightly dependent on the dot energy, with a value $\approx 2.4$, in near agreement with the theoretical ratio.  The agreement is best in the middle of the Coulomb valley and becomes worse closer to the charging peaks, possibly due to the role of higher-excited states not included in our calculations.

The authors thank A. Maharjan and M. Torabi for their help with the low noise circuit construction and J. Markus, M. Ankenbauer and R. Schrott for the technical assistance. T.-M. L. acknowledges SET fabrication support from the Institute for Nanoscale Science and Technology at University of Cincinnati. The research is supported by the NSF DMR (0804199), MWN (07010581, 1108285), and PIRE (0730257), and by the University of Cincinnati.


\begin{thebibliography}{27}
\bibitem{FOOT}  We have also included  averaging of the theoretical conductance
curves to mimic the lock-in process used in collecting the data.  Accordingly, the calculated conductance
is averaged by $\bar{G}=(1/T)\int_0^{T} G(V+A_0\sin(2\pi t/T))dt$, where $A_0$ is the amplitude of the ac-drive of period $T$; 
see caption Fig.\ 1 for parameter values.
\end{thebibliography}
\end{document}